\documentclass[aps,eqsecnum,preprint,floats,epsf,epsfig,nofootinbib,letter]{revtex4}
\usepackage{epsfig}
\usepackage{amsmath}

\begin{document}
\def\be{\begin{eqnarray}}
\def\en{\end{eqnarray}}
\def\non{\nonumber}
\def\ov{\overline}
\def\la{\langle}
\def\ra{\rangle}
\def\half{{{1\over 2}}}
\def\B{{\cal B}}
\def\pr{{\sl Phys. Rev.}~}
\def\prl{{\sl Phys. Rev. Lett.}~}
\def\pl{{\sl Phys. Lett.}~}
\def\np{{\sl Nucl. Phys.}~}
\def\zp{{\sl Z. Phys.}~}
\def\up{\uparrow}
\def\dw{\downarrow}
\def\lsim{ {\ \lower-1.2pt\vbox{\hbox{\rlap{$<$}\lower5pt\vbox{\hbox{$\sim$}
}}}\ } }
\def\gsim{ {\ \lower-1.2pt\vbox{\hbox{\rlap{$>$}\lower5pt\vbox{\hbox{$\sim$}
}}}\ } }

\font\el=cmbx10 scaled \magstep2{\obeylines\hfill March, 2018}

\vskip 1.5 cm

\centerline{\large\bf Singly Cabibbo-suppressed hadronic decays of $\Lambda_c^+$ }

\bigskip
\bigskip
\centerline{\bf Hai-Yang Cheng$^1$, Xian-Wei Kang$^{1, 2}$, Fanrong Xu$^3$}
\medskip
\centerline{$^1$Institute of Physics, Academia Sinica}
\centerline{Taipei, Taiwan 115, Republic of China}
\medskip
\medskip
\centerline{$^2$College of Nuclear Science and Technology, Beijing Normal University}
\centerline{Beijing 100875, People's Republic of China}
\medskip
\medskip
\centerline{$^3$ Department of Physics, Jinan University} \centerline{Guangzhou 510632, People's Republic of China}
\medskip

\bigskip
\bigskip
\centerline{\bf Abstract}
\bigskip
\small
We study singly Cabibbo-suppressed two-body hadronic decays of the charmed baryon $\Lambda_c^+$, namely, $\Lambda_c^+\to \Lambda K^+, p\pi^0, p\eta, n\pi^+,\Sigma^0K^+,\Sigma^+ K^0$. We use the measured rate of  $\Lambda_c^+\to p\phi$ to fix the effective Wilson coefficient $a_2$ for naive color-suppressed modes and the effective number of color $N_c^{\rm eff}$. We rely on the current-algebra approach to evaluate  $W$-exchange and nonfactorizable internal $W$-emission amplitudes,  that is,  the commutator terms for the $S$-wave and the  pole terms for the $P$-wave. Our prediction for $\Lambda_c^+\to p\eta$  is in excellent agreement with the BESIII measurement. The $p\eta$ ($p\pi^0$) mode has a large (small) rate because of a large constructive (destructive) interference between the factorizable and nonfactorizable amplitudes for both $S$- and $P$-waves.
Some of the SU(3) relations such as  $M(\Lambda_c^+\to n\pi^+)=\sqrt{2}M(\Lambda_c^+\to p\pi^0)$ derived under the assumption of sextet dominance are not valid for decays with factorizable terms. Our calculation indicates that  the branching fraction of $\Lambda_c^+\to n\pi^+$  is about 3.5 times larger than that of $\Lambda_c^+\to p\pi^0$.  Decay asymmetries are found to be negative  for all singly Cabibbo-suppressed modes and range from $-0.56$ to $-0.96$.

\pagebreak

\section{Introduction}
The study of hadronic decays of charmed baryons is an old subject (for a review, see \cite{Cheng:2009,Cheng:2015}). For a long time, both experimental and theoretical progresses in this arena were very slow. Almost all the model calculations of two-body nonleptonic decays of charmed baryons were done before millennium and most of the experimental measurements were older ones. Theoretical interest in hadronic weak decays of charmed baryons peaked around the early 1990s and then faded away. To date, we still do not have a good and reliable phenomenological model, not mentioning the QCD-inspired approach as in heavy meson decays, to describe the complicated physics of charmed baryon decays.
\footnote{An exception is the heavy-flavor-conserving hadronic decay of the heavy baryon, for example, $\Xi_c\to\Lambda_c \pi$, which can be reliably studied
within the framework that incorporates both heavy-quark and chiral symmetries \cite{ChengHFC}.}

From the theoretical point of view, baryons being made out of three quarks, in contrast to two quarks for mesons, bring along several essential complications. First of all, the factorization approximation that the hadronic matrix element is factorized into the product of two matrix elements of single currents and that the nonfactorizable term such as the $W$-exchange contribution is negligible relative to the factorizable one is known empirically to be working reasonably well for describing the nonleptonic weak decays of heavy mesons. However, this approximation is a priori not directly applicable to the charmed baryon case as $W$-exchange there, manifested as pole diagrams, is no longer subject to helicity and color suppression. This is different from the naive color suppression of internal $W$-emission. It is known in the heavy meson case that nonfactorizable contributions will render the color suppression of internal $W$-emission ineffective. However, the $W$-exchange in baryon decays is not subject to color suppression even in the absence of nonfactorizable terms.  The
experimental measurements of the decays
$\Lambda_c^+\to\Sigma^0\pi^+,~ \Sigma^+\pi^0$ and
$\Lambda^+_c\to\Xi^0K^+$, which do not receive any factorizable
contributions,
\footnote{At first sight, it appears that the decay modes such as $\Lambda_c^+\to\Sigma^0\pi^+,\Sigma^0K^+$ can proceed through the external $W$-emission process. However, the spectator diquark $ud$ of the $\Lambda_c^+$ is antisymmetric in flavor, while the same diquark in $\Sigma^0$ is symmetric in flavor. Hence, the external $W$-emission is prohibited.}
indicate that $W$-exchange and nonfactorizable internal $W$-emission indeed play an essential role in charmed baryon decays.

Recently, there are two major breakthroughs in charmed-baryon experiments in regard to hadronic weak decays. First of all, it is concerned with the absolute branching fraction of $\Lambda_c^+\to pK^-\pi^+$. Experimentally, nearly all the branching fractions of the
$\Lambda_c^+$ were measured relative to the $pK^-\pi^+$ mode.
On the basis of ARGUS and CLEO data, Particle Data Group (PDG)  had made a model-dependent determination of the absolute branching fraction, ${\cal B}(\Lambda_c^+\to pK^-\pi^+)=(5.0\pm1.3)\%$ \cite{PDG:2014}. Recently, Belle reported a value of $(6.84\pm0.24^{+0.21}_{-0.27})\%$ \cite{Zupanc} from the reconstruction of $D^*p\pi$ recoiling against the $\Lambda_c^+$ production in $e^+ e^-$ annihilation. Hence, the uncertainties are much smaller, and, most importantly, this measurement is model independent! More recently, BESIII has also measured this mode directly with the result ${\cal B}(\Lambda_c^+\to pK^-\pi^+)=(5.84\pm0.27\pm0.23)\%$ \cite{BES:pKpi}. Its precision is comparable to the Belle's result. A new average of $(6.35\pm0.33)\%$ for this benchmark mode is quoted by the PDG \cite{PDG}.

Second, in 2015 BESIII has measured the absolute branching fractions for more than a dozen of  decay modes directly for the first time \cite{BES:pKpi}.  Not only the central values are substantially different from the PDG ones (versions before 2016), but also the uncertainties are significantly improved. For example,  $\B(\Lambda_c^+\to \Sigma^+\omega)=(2.7\pm1.0)\%$ quoted in 2014 PDG \cite{PDG:2014} now becomes $(1.74\pm0.21)\%$  in 2016 PDG \cite{PDG} due to the new measurement of BESIII. In other words, all the PDG values before the 2016 version for the branching fractions of charmed baryon decays become obsolete.

The decay amplitude of the charmed baryon generally consists of factorizable and nonfactorizable contributions. The study of nonfactorizable effects arising from $W$-exchange and internal $W$-emission conventionally relies on the pole model. Under the pole approximation, one usually concentrates on the most important low-lying $1/2^+$ and $1/2^-$ pole states. Consider the charmed baryon decay with a pseudoscalar meson in the final state, $\B_c\to \B+P$. In general, its nonfactorizable $S$- and
$P$-wave amplitudes  are dominated by ${1\over 2}^-$ low-lying baryon resonances and ${1\over 2}^+$ ground-state baryon poles, respectively. It is known that the pole model is reduced to current algebra in the soft pseudoscalar-meson limit. The great advantage of current algebra is that the evaluation of the $S$-wave amplitude does not require the information of the troublesome negative-parity baryon resonances which are not well understood in the quark model. Nevertheless, the use of the pole model is very general and is not limited to the soft meson limit and to the pseudoscalar-meson final state. For example, current algebra is not applicable to the decays  $\B_c\to \B+V$. However,
the estimation of pole amplitudes is a difficult and nontrivial
task since it involves weak baryon matrix elements and strong
coupling constants of ${1\over 2}^+$ and ${1\over 2}^-$ baryon
states.  As a consequence, the evaluation of pole diagrams is far more uncertain
than the factorizable terms.

\begin{table}[tbp]
\footnotesize{
 \caption{Branching fractions (upper entry) and up-down decay asymmetries $\alpha$ (lower entry) of Cabibbo-allowed $\Lambda_c^+\to{\cal
B}+P$ decays in various models. Model results of \cite{Korner,Xu:1992,Cheng:1993,Zen}
have been normalized using the current world average of $\tau(\Lambda_c^+)$ \cite{PDG}. Branching fractions cited from
\cite{Verma98} are for $\phi_{\eta-\eta'}=-23^\circ$ and
$r\equiv |\psi^{{\cal B}_c}(0)|^2/|\psi^{\cal B}(0)|^2=1.4$\,.
 } \label{tab:CF}
\centering
\begin{ruledtabular}
\begin{tabular}{l c c c c c c c c}
~~~~Decay & K\"{o}rner,           & Xu,
& Cheng,             & \,\,\, Ivanov et al \, & \.Zenczykowski & Sharma,  & Expt. \\
        & Kr\"{a}mer \cite{Korner} & Kamal \cite{Xu:1992}
& Tseng \cite{Cheng:1993} & \cite{Ivanov98} & \cite{Zen} & Verma \cite{Verma98}& \cite{PDG} \\
& & & CA \quad Pole & & &  & \\
 \hline
$\Lambda^+_c\to \Lambda \pi^+$ & input & 1.62 & 1.46 \quad 0.88 & 0.79 & 0.52 & 1.12 & $1.30\pm 0.07$ \\
$\Lambda^+_c\to p \bar K^0$ & input & 1.20 & 3.64 \quad 1.26 & 2.06 & 1.71 & 1.64 & $3.16\pm0.16$ \\
$\Lambda^+_c\to \Sigma^0 \pi^+$ & 0.32 & 0.34 & 1.76 \quad 0.72 & 0.88 & 0.39 & 1.34 & $1.29\pm0.07$ \\
$\Lambda^+_c\to \Sigma^+ \pi^0$ & 0.32 & 0.34 & 1.76 \quad 0.72 & 0.88 & 0.39 & 1.34 & $1.24\pm0.10$ \\
$\Lambda^+_c\to \Sigma^+ \eta$ & 0.16 & & & 0.11 & 0.90 & 0.57 & $0.70\pm0.23$ \\
$\Lambda^+_c\to \Sigma^+ \eta^\prime$ & 1.28 & & & 0.12& 0.11 &  0.10 &  \\
$\Lambda^+_c\to \Xi^0 K^+$ & 0.26 & 0.10 & & 0.31 & 0.34 &  0.13 & $0.50\pm0.12$ \\
\hline
$\Lambda^+_c\to \Lambda \pi^+$ & $-$0.70 & $-$0.67 & $-$0.99 \quad
$-$0.95 &  $-$0.95 & $-$0.99 &  $-$0.99 & $-$0.91 $\pm$ 0.15\\
$\Lambda^+_c\to p \bar K^0$ & $-$1.0 & 0.51 & $-$0.90 \quad $-$0.49 & $-$0.97 & $-$0.66 & $-$0.99 &  \\
$\Lambda^+_c\to \Sigma^0 \pi^+$ & 0.70 & 0.92 & $-$0.49 \quad~ 0.78 & 0.43 & 0.39 & $-$0.31 & \\
$\Lambda^+_c\to \Sigma^+ \pi^0$ & 0.70 & 0.92 &  $-$0.49 \quad~ 0.78 & 0.43
& 0.39 & $-$0.31  & $-$0.45$\pm$ 0.32   \\
$\Lambda^+_c\to \Sigma^+ \eta$ & 0.33 & & & 0.55 & 0 & $-$0.91 &  \\
$\Lambda^+_c\to \Sigma^+ \eta^\prime$ & $-$0.45 & & & $-$0.05 & $-$0.91 & 0.78 &  \\
$\Lambda^+_c\to \Xi^0 K^+$ & 0 & 0 & & 0 & 0 & 0  & \\
\end{tabular}
\end{ruledtabular}
 }
\end{table}

In Table \ref{tab:CF} we show various model calculations of branching fractions  and up-down decay asymmetries  of Cabibbo-allowed $\Lambda_c^+\to{\cal B}+P$ decays. Two explicit pole-model calculations were carried out in \cite{Xu:1992} and \cite{Cheng:1992,Cheng:1993} and a variant of the pole model was
considered in \cite{Zen}. In \cite{Verma98}, the $S$-wave amplitude was calculated using current algebra. Similar calculations based on current algebra also can be found in \cite{Cheng:1993} (denoted by CA in Table \ref{tab:CF}).
Authors of \cite{Korner} chose to use the covariant quark model to tackle the three-body transition amplitudes (rather than two-body transitions) directly. This work was further developed in \cite{Ivanov98}. We see from Table \ref{tab:CF} that the predicted rates by most of the models except current algebra are generally below experiment. Moreover, the pole model, the covariant quark model and its variant all predict a positive decay asymmetry $\alpha$ for both $\Lambda_c^+\to \Sigma^+\pi^0$ and $\Sigma^0\pi^+$, while it is measured to be $-0.45\pm0.31\pm0.06$ for $\Sigma^+\pi^0$ by CLEO \cite{CLEO:alpha}. In contrast, current algebra always leads to a negative decay asymmetry for aforementioned two modes: $-0.49$ in \cite{Cheng:1993}, $-0.31$ in \cite{Verma98}, $-0.76$ in \cite{Zen:1993} and $-0.47$ in \cite{Datta}.  BESIII will measure decay asymmetry parameters for $\Lambda_c^+\to \Lambda\pi^+,\Sigma^0\pi^+,\Sigma^+\pi^0$ and $p\bar K^0$ and the sensitivity for measuring $\alpha_{\Sigma^+\pi^0}$ is estimated to be $(10\sim 77)\%$ \cite{BES:alpha}. It will be of great interest to see if the negative sign of $\alpha_{\Sigma^+\pi^0}$ measured by CLEO is confirmed.

Writing the nonfactorizable $S$-wave amplitude as
\be \label{eq:onshell}
A=A^{\rm CA}+(A-A^{\rm CA}),
\en
the term $(A-A^{\rm CA})$ can be regarded as an on-shell correction to the current-algebra result. It turns out that in the existing pole model calculations
\cite{Cheng:1992,Cheng:1993,Xu:1992}, the on-shell correction  $(A-A^{\rm CA})$ always has a sign opposite to that of $A^{\rm CA}$. Moreover, its magnitude is sometimes even bigger than $|A^{\rm CA}|$ for some of the decays such as $\Lambda_c^+\to \Sigma^0\pi^+,\Sigma^+\pi^0$. That is, the on-shell correction is large enough to flip the sign of the parity-violating (PV) amplitudes.
This explains  the smaller calculated rate in the pole model and the sign difference of $\alpha_{\Sigma^+\pi^0,\Sigma^0\pi^+}$ between the pole model and current algebra. If the negative sign of $\alpha_{\Sigma^+\pi^0}$ is confirmed, this means that the on-shell correction $(A-A^{\rm CA})$ has been overestimated in previous pole model calculations probably owing to our poor knowledge of the negative-parity baryon resonances. The empiric fact that current algebra seems to work reasonably well for $\Lambda_c^+\to \B+P$ is a bit surprising and annoying since the pseudoscalar meson produced in $\Lambda_c^+$ decays is generally far from being soft. We plan to examine this important issue and the pole model in a separate work.

In this work we will focus on singly Cabibbo-suppressed hadronic decays of the $\Lambda_c^+$, specifically, $\Lambda_c^+\to \Lambda K^+, p\pi^0, p\eta, n\pi^+,\Sigma^0K^+,\Sigma^+ K^0$. Among them, evidence of $\Lambda_c^+\to p\eta$
was found by BESIII recently \cite{BES:peta}, while a stringent upper limit on $\Lambda_c^+\to p\pi^0$ was also set. Besides dynamical model calculations, two-body nonleptonic decays of charmed baryons have
been analyzed in terms of SU(3)-irreducible-representation
amplitudes \cite{Savage,Verma}. However, the quark-diagram scheme
(i.e., analyzing the decays in terms of topological quark-diagram amplitudes)
has the advantage that it is more intuitive and easier for implementing
model calculations.  A general formulation of the quark-diagram scheme for charmed baryons is given in \cite{Chau} (see also \cite{Kohara}).  Analysis of  Cabibbo-suppressed decays using SU(3) flavor symmetry was first carried out in \cite{Sharma}. This approach became popular recently \cite{Lu,Geng:Lambdac,Geng:2017mxn,Geng:2018}. Nevertheless, we shall perform dynamical model calculations based on current algebra.

This work is organized as follows. In Sec. II we set up the formalism for analyzing factorizable and nonfactorizable contributions to singly Cabibbo-suppressed decays of the charmed baryon $\Lambda_c^+$. Numerical model calculations and discussions are presented in Sec. III.
Sec. IV gives our conclusion. Appendix A is devoted to the study of the decay $\Lambda_c^+\to p\phi$ to fix the relevant Wilson coefficient. The MIT bag model evaluation of baryon matrix elements is sketched in Appendix B. Axial-vector form factors and baryon wave functions relevant to the present work are summarized in Appendices C and D, respectively.

\section{Formalism}
The effective weak Hamiltonian for singly Cabibbo-suppressed decays at the scale $\mu=m_c$ reads \cite{Buchalla}
\be  \label{eq:H}
H_{\rm eff}={G_F\over \sqrt{2}}\sum_{q=d,s}V^*_{uq}V_{cq}(c_1O_1^q+c_2O_2^q)+h.c.,
\en
with $q=d,s$ and the four-quark operators are given by
\be
O_1^q=(\bar qc)(\bar uq), \qquad O_2^q=(\bar qq)(\bar uc),
\en
with $(\bar q_1q_2)\equiv \bar q_1\gamma_\mu(1-\gamma_5)q_2$. For the Wilson coefficients, we shall use the lowest order values $c_1=1.346$ and $c_2=-0.636$ obtained at the scale $\mu=1.25$ GeV with $\Lambda^{(4)}_{\ov {\rm MS}}=325$ MeV (see Tables VI and VII of \cite{Buchalla}). Because in this work we will not consider effects of CP violation, we shall assume real CKM matrix elements for simplicity thereafter.

The general amplitude for $\B_i\to \B_f+P$ is given by
\begin{eqnarray} \label{eq:Amp}
M(\B_i\to \B_f+P)=i\bar u_f(A-B\gamma_5)u_i,
\end{eqnarray}
where $A$ and $B$ are the $S$- and $P$-wave amplitudes, respectively. Note that
if we write $M(\B_i\to \B_f+P)=i\bar u_f(A+B\gamma_5)u_i$, the $P$-wave amplitudes given in Eqs. (\ref{eq:facamp}) and (\ref{eq:amppole}) below and the decay asymmetry $\alpha$  in Eq. (\ref{eq:alpha}) will be flipped in sign. The decay amplitude
generally consists of factorizable and nonfactorizable ones
\be
M(\B_i\to \B_f+P)=M(\B_i\to \B_f+P)^{\rm fac}+M(\B_i\to \B_f+P)^{\rm nf}.
\en
While the factorizable amplitude vanishes in the soft meson limit, the nonfactorizable one is not.

\subsection{Factorizable contributions}
We first consider the factorizable amplitudes for some of singly Cabibbo-suppressed modes:
\be
M(\Lambda_c^+\to \Lambda K^+)^{\rm fac} &=& {G_F\over \sqrt{2}}V_{cs}V_{us}a_1\la K^+|(\bar us)|0\ra\la\Lambda|(\bar s c)|\Lambda_c^+\ra, \non \\
M(\Lambda_c^+\to p\pi^0)^{\rm fac} &=& {G_F\over \sqrt{2}}V_{cd}V_{ud}a_2\la \pi^0|(\bar dd)|0\ra\la p|(\bar u c)|\Lambda_c^+\ra,
\en
where $a_1=c_1+{c_2\over N_c}$ for
the external (color-allowed) $W$-emission amplitude and $a_2=c_2+{c_1\over N_c}$ for internal (color-suppressed) $W$-emission in nave factorization.
In terms of the decay constants and form factors defined by
\footnote{There is a sign ambiguity for the one-body matrix element. We define Eq. (\ref{eq:1bodym.e.}) in such a way that a correct relative sign between the factorizable and nonfactorizable amplitudes, e.g. between Eqs. (\ref{eq:fac}) and (\ref{eq:amppole}), is ensured.}
\be \label{eq:1bodym.e.}
\la K^+(q)|(\bar us)|0\ra=-if_K q_\mu, \qquad \la \pi^0(q)|(\bar dd)|0\ra={i\over\sqrt{2}}f_\pi q_\mu,
\en
and
\be \label{eq:FF}
\la \Lambda(p_\Lambda)|(\bar sc)|\Lambda_c^+(p_{\Lambda_c})\ra &=& \bar u_{\Lambda}\Bigg[f_1^{\Lambda_c\Lambda}(q^2)\gamma_\mu-f_2^{\Lambda_c\Lambda}(q^2)i\sigma_{\mu\nu}{q^\nu\over m_{\Lambda_c}}+
f_3^{\Lambda_c\Lambda}(q^2){q_\mu\over m_{\Lambda_c}} \non \\
&& -\left( g_1^{\Lambda_c\Lambda}(q^2)\gamma_\mu-g_2^{\Lambda_c\Lambda}(q^2)i\sigma_{\mu\nu}{q^\nu\over m_{\Lambda_c}}
+g_3^{\Lambda_c\Lambda}(q^2) {q_\mu\over m_{\Lambda_c}}\right)\gamma_5 \Bigg]u_{\Lambda_c},
\en
with $q=p_{\Lambda_c}-p_\Lambda$, we obtain
\be \label{eq:fac}
M^{\rm fac}(\Lambda_c^+\to \Lambda K^+) &=& -i{G_F\over\sqrt{2}}V_{cs}V_{us}\,a_1f_K \left[(m_{\Lambda_c}-m_{\Lambda})f_1^{\Lambda_c\Lambda}(m_K^2)
 +(m_{\Lambda_c}+m_{\Lambda})g_1^{\Lambda_c\Lambda}(m_K^2)\gamma_5\right], \non \\
M^{\rm fac}(\Lambda_c^+\to p\pi^0) &=& i{G_F\over 2}V_{cd}V_{ud}\,a_2f_\pi \left[(m_{\Lambda_c}-m_{p})f_1^{\Lambda_c p}(m_\pi^2)
 +(m_{\Lambda_c}+m_{p})g_1^{\Lambda_c p}(m_\pi^2)\gamma_5\right],
\en
where we have neglected contributions from the form factors $f_3$ and $g_3$.
We have learned
from charmed meson decays that naive factorization
does not work for color-suppressed decay modes. Empirically,
it was realized in the 1980s that if the Fierz-transformed terms
characterized by $1/N_c$ are dropped, the discrepancy between
theory and experiment will be greatly improved \cite{Fuk}. This leads
to the so-called large-$N_c$ approach for describing hadronic $D$
decays \cite{Buras}.
As the discrepancy between theory and experiment for charmed meson
decays gets much improved in the $1/N_c$ expansion method, it is
natural to ask if this scenario also works in the baryon sector.
This issue can be settled down by the experimental measurement of
the Cabibbo-suppressed mode $\Lambda_c^+\to p\phi$, which receives
contributions only from the factorizable diagrams \cite{Cheng:1992}.
Using the recent BESIII measurement of $\Lambda_c^+\to p\phi$ \cite{BES:pphi}, we obtain $|a_2|=0.45\pm0.03$, corresponding to $N_c^{\rm eff}\approx 7$ (see Appendix A below). Recall that $a_2=-0.19$ for $N_c=3$. Hence, color suppression in the factorizable amplitude is not operative.

For $\Lambda_c^+\to p\eta^{(')}$ decays, we need to consider the $\eta-\eta'$ mixing parametrized by
\begin{eqnarray}
|\eta\rangle &=& \cos\phi|\eta_q\rangle-\sin\phi|\eta_s\rangle, \non \\
|\eta'\rangle &=& \sin\phi|\eta_q\rangle+\cos\phi|\eta_s\rangle,
\end{eqnarray}
where the flavor states $q\bar q=(u\bar u+d\bar d)/\sqrt{2}$ and $s\bar s$ are labeled as $\eta_q$ and $\eta_s$, respectively.
The mixing angle $\phi$ is determined to be $39.3^\circ\pm 1.0^\circ$ in the
Feldmann-Kroll-Stech mixing scheme \cite{Kroll},  which is consistent
with the recent result $\phi=42^\circ\pm2.8^\circ$ extracted from
the CLEO data \cite{Hietala}. The factorizable amplitudes then read
\be \label{eq:facamp}
A^{\rm fac}(\Lambda_c^+\to p\eta^{(')}) &=& -{G_F\over \sqrt{2}}\,a_2 \left(V_{cs}V_{us}f_{\eta^{(')}}^s+{1\over\sqrt{2}}V_{cd}V_{ud}f_{\eta^{(')}}^q\right) (m_{\Lambda_c}-m_{p})f_1^{\Lambda_c p}(m_\eta^2),  \non \\
B^{\rm fac}(\Lambda_c^+\to p\eta^{(')}) &=& {G_F\over \sqrt{2}}\,a_2 \left(V_{cs}V_{us}f_{\eta^{(')}}^s+{1\over\sqrt{2}}V_{cd}V_{ud}f_{\eta^{(')}}^q\right) (m_{\Lambda_c}+m_{p})g_1^{\Lambda_c p}(m_\eta^2),
\en
where the decay constants are defined by
\be
\la \eta^{(')}|\bar q\gamma_\mu\gamma_5 q|0\ra=i{1\over \sqrt{2}}f_{\eta^{(')}}^q q_\mu, \qquad
\la \eta^{(')}|\bar s\gamma_\mu\gamma_5 s|0\ra=if_{\eta^{(')}}^s q_\mu.
\en
We follow \cite{Kroll} to use
\be \label{eq:decayconst}
f_\eta^q= 107\, {\rm MeV}, \qquad f^s_\eta= -112\, {\rm MeV}, \qquad
f_{\eta'}^q= 89\, {\rm MeV},\qquad f^s_{\eta'}=137\, {\rm MeV}
\en
for $\phi=39.3^\circ$.

\subsection{Nonfactorizable contributions}

Besides factorizable terms, there exist nonfactorizable contributions arising from $W$-exchange (see e.g. diagrams $E_{1,2,3}$ in Fig. \ref{fig:SCS} below) or nonfactorizable internal $W$-emission (e.g. diagram $C_2$ in Fig. \ref{fig:SCS}). How do we tackle with the nonfactorizable contributions? One popular approach is to consider the contributions from all possible intermediate states. Among all possible pole contributions, including resonances and continuum states, one usually focuses on the most important poles such as the low-lying $1/2^+$ and $1/2^-$ states, known as pole approximation. More specifically, the $S$-wave amplitude is dominated by the low-lying $1/2^-$ resonances and the $P$-wave one governed by the ground-state  poles. The nonfactorizable $S$- and $P$-wave amplitudes for the process $\B_i\to \B_f+M$ are then given by  \cite{Cheng:1992}
\be \label{eq:amppole}
A^{\rm pole} &=& -\sum_{\B_n^*(1/2^-)}\left[ {g_{_{\B_f\B_{n^*}M}b_{n^*i}}\over m_i-m_{n^*}}+ {b_{fn^*}g_{_{\B_{n^*}\B_iM}}\over m_f-m_{n^*}}\right]+\cdots, \non \\
B^{\rm pole} &=& -\sum_{\B_n}\left[ {g_{_{\B_f\B_nM}} a_{ni}\over m_i-m_{n}}+ {a_{fn}g_{_{\B_{n}\B_iM}}\over m_f-m_{n}}\right]+\cdots
\en
respectively. Ellipses in the above equation denote other pole contributions which are negligible for our purposes,
\footnote{For example, contributions to the $S$-wave amplitude from the parity-violating matrix elements $b_{ij}$ defined
in Eq. (\ref{eq:a&b}) are much smaller than the parity-conserving ones $a_{ij}$, which have been shown explicitly in \cite{Ebert,Cheng:1985}.}
and the baryon-baryon matrix elements are defined by \cite{Cheng:1992}
\be \label{eq:a&b}
\la \B_i|H_{\rm eff}|\B_j\ra=\bar u_i(a_{ij}-b_{ij}\gamma_5)u_j, \qquad
\la \B_i^*(1/2^-)|H_{\rm eff}^{\rm PV}|\B_j\ra = ib_{i^*j}\bar u_i u_j.
\en
When $M=P$, one can apply the Goldberger-Treiman relation for the strong coupling $g_{_{\B'\B P}}$ and its generalization for $g_{_{\B^*\B P}}$
\be \label{eq:GT}
g_{_{\B'\B P^a}}= {\sqrt{2}\over f_{P^a}}(m_{\B'}+m_\B)g^A_{\B'\B}, \qquad
g_{_{\B^*\B P^a}}= {\sqrt{2}\over f_{P^a}}(m_{\B^*}-m_\B)g^A_{\B^*\B},
\en
to express Eq. (\ref{eq:amppole}) as
\be
A^{\rm pole} &=& - {\sqrt{2}\over f_{P^a}} \sum_{\B_n^*(1/2^-)} \left[ g_{\B_f\B_{n^*}}^A {p\!\!\!/_f -m_{n^*}\over p\!\!\!/_i-m_{n^*}}b_{n^*i} - b_{fn^*} {p\!\!\!/_i -m_{n^*}\over p\!\!\!/_f-m_{n^*}} g_{\B_{n^*}\B_i}^A  \right], \non \\
B^{\rm pole} &=& -{\sqrt{2}\over f_{P^a}} \sum_{\B_n} \left[ g_{\B_f\B_n}^A {m_f+m_n\over m_i-m_n} a_{ni}+ a_{fn}{m_i+m_n\over m_f-m_n} g_{\B_{n}\B_i}^A\right],
\en
with the decay constant normalized to $f_{P^3}=f_\pi=132$ MeV.
In the soft pseudoscalar-meson limit, $p_f=p_i$ and hence the $S$-wave amplitude can be recast to the form
\be
A^{\rm com} &=&  - {\sqrt{2}\over f_{P^a}} \sum_{\B_n^*(1/2^-)} \left[ \la \B_f|Q_5^a|\B_n^*\ra \la \B_n^*|H_{\rm eff}^{\rm PV}|\B_i\ra-
\la \B_f|H_{\rm eff}^{\rm PV}|\B_n^*\ra \la \B_n^*|Q_5^a|\B_i\ra
\right]  \non \\
&=& -{\sqrt{2}\over f_{P^a}}\la \B_f| [Q_5^a, H_{\rm eff}^{\rm PV}]|\B_i\ra,
\en
with
\be
Q^a=\int d^3x\,\bar q\gamma_0{\lambda^a\over 2}q,  \qquad Q^a_5=\int d^3x\,\bar q\gamma_0\gamma_5{\lambda^a\over 2}q.
\en
The above expression for $A^{\rm com}$  is precisely the well-known soft-pion theorem in the current-algebra approach. Using the relation $[Q_5^a, H_{\rm eff}^{\rm PV}]=[Q^a, H_{\rm eff}^{\rm PC}]$, we see that in the soft meson limit, the parity-violating amplitude is reduced to a simple commutator term expressed in terms of parity-conserving matrix elements. Therefore, the great advantage of current algebra is that the evaluation of the parity-violating $S$-wave amplitude does not require the information of the negative-parity $1/2^-$ poles.

To apply the soft-meson theorem, we notice that
\be \label{eq:soft}
&& A^{\rm com}(\B_i\to \B_f\pi^0)=-{\sqrt{2}\over f_\pi}\la \B_f| [I_3, H_{\rm eff}^{\rm PC}]|\B_i\ra, \qquad A^{\rm com}(\B_i\to \B_f\pi^\pm)=-{1\over f_\pi}\la \B_f| [I_\mp, H_{\rm eff}^{\rm PC}]|\B_i\ra,  \non \\
&& A^{\rm com}(\B_i\to \B_f K^\pm)=-{1\over f_K}\la \B_f| [V_\mp, H_{\rm eff}^{\rm PC}]|\B_i\ra,  \quad A^{\rm com}(\B_i\to \B_f \stackrel{(-)}{K^0})=-{1\over f_\pi}\la \B_f| [U_\mp, H_{\rm eff}^{\rm PC}]|\B_i\ra, \non \\
&& A^{\rm com}(\B_i\to \B_f \eta_8)= -\sqrt{3\over 2}{1\over f_{\eta_8}}\la \B_f|[Y, H_{\rm eff}^{\rm PC}]|\B_i\ra,
\en
where $I_\pm,U_\pm$ and $V_\pm$ are isospin, $U$-spin and $V$-spin ladder operators, respectively, with
\be \label{eq:IUV}
I_+|d\ra=|u\ra, \quad I_-|u\ra=|d\ra, \quad U_+|s\ra=|d\ra, \quad U_-|d\ra=|s\ra, \quad V_+|s\ra=|u\ra, \quad V_-|u\ra=|s\ra.
\en
The use of the hypercharge $Y={2\over\sqrt{3}}Q^8$ has been made in the last line of  Eq. (\ref{eq:soft}). In the SU(3) case, the hypercharge is given by the well-known relation $Y=B+S$. However, its generalization to the SU(4) case depends on the generalized definition of the hypercharge. For example, $Y=B+S-C$  is derived in the textbook of \cite{Lichtenberg}, while the relation $Y=B+S+C$ also can be found in the literature. For our purpose, we will adopt the first one, so that $Y(p)=1$ and $Y(\Lambda_c^+)=0$. We will come back to this point in Sec. III.

Applying Eq. (\ref{eq:IUV}) to the commutator terms for singly Cabibbo-suppressed modes:
$\Lambda_c^+\to \Lambda K^+, p\pi^0, p\eta, n\pi^+,\Sigma^0K^+,\Sigma^+ K^0$, we obtain
\be \label{eq:Swave}
&& A^{\rm com}(\Lambda_c^+\to \Lambda K^+) = {1\over f_K}\left(\sqrt{3/2}\,a_{p\Lambda_c}+a_{\Lambda\Xi_c^0}\right), \qquad
A^{\rm com}(\Lambda_c^+\to p\pi^0) = -{1\over \sqrt{2}f_\pi}a_{p\Lambda_c}, \non \\
&& A^{\rm com}(\Lambda_c^+\to \Sigma^0 K^+) = {1\over \sqrt{2}f_K}\left(a_{p\Lambda_c}+\sqrt{2}a_{\Sigma^0\Xi_c^0 }\right), \quad A^{\rm com}(\Lambda_c^+\to n \pi^+) = -{1\over f_\pi}a_{p\Lambda_c},
 \\
&& A^{\rm com}(\Lambda_c^+\to \Sigma^+ K^0) = {1\over f_K}\left(a_{p\Lambda_c}-a_{\Sigma^+\Xi_c^+ }\right), \qquad \quad~
A^{\rm com}(\Lambda_c^+\to p\eta_8) =  -\sqrt{3\over 2}{1\over f_{\eta_8}}a_{p\Lambda_c}, \non
\en
for $S$-wave amplitudes with $a_{_{\B \B_c}}\equiv \la \B|H_{\rm eff}^{^{\rm PC}}|\B_c\ra$. For $P$-wave amplitudes, we have
\be \label{eq:Pwave}
B^{\rm ca}(\Lambda_c^+\to p\pi^0) &=& -{\sqrt{2}\over f_\pi}\left(g_{pp}^{A(\pi^0)}{m_p+m_p\over m_{\Lambda_c}-m_p} a_{p\Lambda_c}+ a_{p\Sigma_c^+}{m_{\Lambda_c}+m_{\Sigma_c}\over m_p-m_{\Sigma_c}} g_{\Sigma_c^+\Lambda_c}^{A(\pi^0)} + a_{p\Lambda_c}{m_{\Lambda_c}+m_{\Lambda_c}\over m_p-m_{\Lambda_c}} g_{\Lambda_c\Lambda_c}^{A(\pi^0)}\right), \non \\
B^{\rm ca}(\Lambda_c^+\to n \pi^+) &=& -{1\over f_\pi}\left(g_{np}^A{m_n+m_p\over m_{\Lambda_c}-m_p} a_{p\Lambda_c}+ a_{n\Sigma_c^0}{m_{\Lambda_c}+m_{\Sigma_c}\over m_n-m_{\Sigma_c}} g_{\Sigma_c^0\Lambda_c}^A \right),  \non \\
B^{\rm ca}(\Lambda_c^+\to p\eta_8) &=& -{\sqrt{2}\over f_{\eta_8}}\left(g_{pp}^{A(\eta_8)}{m_p+m_p\over m_{\Lambda_c}-m_p} a_{p\Lambda_c}+ a_{p\Sigma_c^+}{m_{\Lambda_c}+m_{\Sigma_c}\over m_p-m_{\Sigma_c}} g_{\Sigma_c^+\Lambda_c}^{A(\eta_8)} + a_{p\Lambda_c}{m_{\Lambda_c}+m_{\Lambda_c}\over m_p-m_{\Lambda_c}} g_{\Lambda_c\Lambda_c}^{A(\eta_8)}\right), \non  \\
B^{\rm ca}(\Lambda_c^+\to \Lambda K^+) &=& -{1\over f_K}\left(g_{\Lambda p}^A{m_\Lambda+m_p\over m_{\Lambda_c}-m_p} a_{p\Lambda_c }+ a_{\Lambda\Xi_c^0}{m_{\Lambda_c}+m_{\Xi_c}\over m_\Lambda-m_{\Xi_c}} g_{\Xi_c^{0}\Lambda_c}^A + a_{\Lambda\Xi_c^{'0}}{m_{\Lambda_c}+m_{\Xi_c'}\over m_\Lambda-m_{\Xi_c'}} g_{\Xi_c^{'0}\Lambda_c}^A\right), \non \\
B^{\rm ca}(\Lambda_c^+\to \Sigma^0 K^+) &=&  -{1\over f_K}\Big(g_{\Sigma^0 p}^A{m_\Sigma+m_p\over m_{\Lambda_c}-m_p} a_{p\Lambda_c}+ a_{\Sigma^0\Xi_c^0 }{m_{\Lambda_c}+m_{\Xi_c} \over m_\Sigma-m_{\Xi_c}} g_{\Xi_c^0\Lambda_c}^A  \\
  && \qquad +  a_{\Sigma^0\Xi_c^{'0}}{m_{\Lambda_c}+m_{\Xi_c'} \over m_\Sigma-m_{\Xi_c'}} g_{\Xi_c^{'0}\Lambda_c}^A\Big), \non \\
B^{\rm ca}(\Lambda_c^+\to \Sigma^+ K^0) &=&  -{1\over f_K}\Big(g_{\Sigma^+p}^A{m_\Sigma+m_p\over m_{\Lambda_c}-m_p} a_{p\Lambda_c}+ a_{\Sigma^+\Xi_c^+ }{m_{\Lambda_c}+m_{\Xi_c}\over m_\Sigma-m_{\Xi_c}} g_{\Xi_c^+\Lambda_c}^A \non \\
&& \qquad +a_{\Sigma^+\Xi_c^{'+}}{m_{\Lambda_c}+m_{\Xi_c'}\over m_\Sigma-m_{\Xi_c'}} g_{\Xi_c^{'+}\Lambda_c}^A \Big), \non
\en
where the superscript $\pi^0$ of $g_{pp}^{A(\pi^0)}$ implies that the form factor $g_{pp}^A$ is evaluated using the the axial-vector current corresponding to $P^3=\pi^0$, and likewise for  the superscript $\eta_8$ of $g_{pp}^{A(\eta_8)}$.
In Eqs. (\ref{eq:soft}) and (\ref{eq:Pwave}), $\eta_8$ is the octet component of the $\eta$ and $\eta'$
\be
\eta=\cos\theta\eta_8-\sin\theta\eta_0, \qquad \eta'=\sin\theta\eta_8+\cos\theta\eta_0,
\en
with $\theta=-15.4^\circ$ \cite{Kroll}. For the singlet component $\eta_0$, the soft pseudoscalar meson theorem is not applicable. Hence, we will not consider the $S$-wave amplitude of $\Lambda_c^+\to p\eta_0$ within the current-algebra framework. As shown in Appendix C, the axial-vector form factor vanishes for antitriplet-antitriplet heavy baryon transitions, i.e. $g_{\B_{\bar 3}\B_{\bar 3}}^A=0$. Hence, in the $P$-wave amplitudes we can drop those terms with $g^A_{\Lambda_c\Lambda_c}$ or $g^A_{\Xi_c\Lambda_c}$.

\subsection{Baryon matrix elements}

To evaluate the nonfactorizable amplitudes we need to know the baryon matrix elements and the axial-vector form factor at $q^2=0$, $g^A_{\B'\B}$.
For the matrix elements, we write
\be
a_{\B\B_c} \equiv  \la \B|H_{\rm eff}^{\rm PC}|\B_c \ra = {G_F\over2\sqrt{2}}\sum_{q=d,s}V_{cq}V_{uq}\la \B|c_+O_+^q+c_-O_-^q|\B_c \ra,
\en
with $O_\pm^q=O_1^q\pm O_2^q=(\bar qc)(\bar uq)\pm(\bar qq)(\bar uc)$ and $c_\pm=c_1\pm c_2$. Since the four-quark operator $O_+$ is symmetric in color indices while $O_-$ is antisymmetric, the former does not contribute to the baryon transition matrix element since the baryon wave function is totally antisymmetric in color. Hence,
\be \label{eq:a}
a_{\B \B_c} = {G_F\over2\sqrt{2}}\sum_{q=d,s}V_{cq}V_{uq}(c_1-c_2)\la \B|O_-^q|\B_c \ra.
\en
We shall evaluate the matrix elements using the MIT bag model (see Appendix B).
The relevant PC matrix elements are
\be \label{eq:m.e.}
&& \la p|O_-^d|\Sigma_c^+\ra={2\sqrt{2}\over 3}(-X_1^d+9X_2^d)(4\pi),  ~~ \la \Sigma^+|O_-^d|\Sigma_c^+\ra=-{2\sqrt{2}\over 3}(-X_1^d+9X_2^d)(4\pi),\non \\
&& \la p|O_-^d|\Lambda_c^+\ra ={4\over \sqrt{6}}(X_1^d+3X_2^d)(4\pi), \qquad \la n|O_-^d|\Sigma_c^0\ra = 5(X_1^d+X_2^d)(4\pi), \non \\
&& \la \Lambda|O_-^d|\Xi_c^0\ra=-4X_2^d(4\pi), \qquad\qquad\qquad \la \Lambda|O_-^s|\Xi_c^0\ra= 2(-X_1^s+X_2^s)(4\pi), \non  \\
&& \la \Lambda|O_-^d|\Xi_c^{'0}\ra=-4\sqrt{3}X_2^d(4\pi), \qquad\quad\quad~~  \la \Lambda|O_-^s|\Xi_c^{'0}\ra=-{2\over\sqrt{3}}(X_1^s+3X_2^s)(4\pi) ,  \\
&&  \la \Sigma^0|O_-^d|\Xi_c^0\ra= -{4\over\sqrt{3}}X_1^d(4\pi), \qquad\qquad~~ \la \Sigma^0|O_-^s|\Xi_c^0\ra= -{2\over\sqrt{3}}(X_1^s+3X_2^s)(4\pi), \non \\
&&  \la \Sigma^0|O_-^d|\Xi_c^{'0}\ra={4\over 3}X_1^d(4\pi), \qquad\qquad\qquad \la \Sigma^0|O_-^s|\Xi_c^{'0}\ra=-{2\over 3}(X_1^s-9X_2^s)(4\pi) , \non \\
&& \la \Sigma^+|O_-^s|\Xi_c^+\ra= 2\sqrt{6}X_2^s(4\pi), \qquad\qquad~~ \la \Sigma^+|O_-^s|\Xi_c^{'+}\ra= {4\over 3\sqrt{2}}(X_1^s-9X_2^s)(4\pi), \non
\en
where $X_1^q$ and $X_2^q$ with $q=d,s$ are the bag integrals defined in Eq. (\ref{eq:baginteg}). The numerical values of the bag integrals can be found in Eq. (\ref{eq:Xnum}). It should be stressed that the relative signs of matrix elements are fixed by the baryon wave functions given in Appendix D.

For the $q^2$ dependence of the form factors defined in Eq. (\ref{eq:FF}), we follow the conventional practice to assume a pole dominance
\be
f_i(q^2)={f_i(0)\over (1-q^2/m_V^2)^n }, \qquad g_i(q^2)={g_i(0)\over (1-q^2/m_A^2)^n },
\en
with $n=2$ or 1,
where $m_V(m_A)$ is the pole mass of the vector (axial-vector) meson with the same quantum number under consideration, for example, $m_V=m_{D^*_s}$ and $m_A=m_{D_{s1}(2536)}$ for $\Lambda_c\to \Lambda$ transition.
Form factors $f_i$ and $g_i$ for $\Lambda_c\to \Lambda$ and $\Lambda_c\to p$ transitions at zero recoil and at maximal recoil $q^2=0$ have been calculated in the literature \cite{Gutsche,Liu:2009sn,Cheng:1992,PerezMarcia}. Presumably, the SU(3) relation
\be
f_i^{\Lambda_c p}(q^2)=-\sqrt{3\over 2}f_i^{\Lambda_c\Lambda}(q^2), \qquad
g_i^{\Lambda_c p}(q^2)=-\sqrt{3\over 2}g_i^{\Lambda_c\Lambda}(q^2),
\en
should be respected at zero recoil $q^2=(m_i-m_f)^2$. For our purpose, we shall follow \cite{Gutsche} to use
\be
f_1^{\Lambda_c p}(0)=-0.470, \qquad g_1^{\Lambda_c p}(0)=-0.414\,.
\en
for $\Lambda_c-p$ transition.
Form factors for $\Lambda_c-\Lambda$ transition will be discussed in Sec. III below.

\begin{table}[tbp]
 \caption{The predicted $S$- and $P$-wave amplitudes of singly Cabibbo-suppressed decays $\Lambda_c^+\to\B+P$ in units of $G_F 10^{-2}{\rm GeV}^2$. Branching fractions and the asymmetry parameter $\alpha$ are shown in the last three columns. Experimental results are taken from \cite{BES:peta,PDG}.
 } \label{tab:S&P}
\centering
\begin{ruledtabular}
\begin{tabular}{lrrrrrr |ccr}
Channel & $A^{\rm fac}$  & $A^{\rm com}$  & $A^{\rm tot}$  & $B^{\rm fac}$ & $B^{\rm ca}$ & $B^{\rm tot}$  & ${\cal B}_{\rm theo}$  & ${\cal B}_{\rm expt}$ & $\alpha_{\rm theo}$   \\
\hline
$\Lambda_c^+\to p\pi^0$ &  $-0.41$ & 0.81 & 0.40 & 0.87 & $-1.57$ & $-0.70$ & $0.75\times 10^{-4}$ & $<2.7\times 10^{-4}$ & $-0.95$ \\
$\Lambda_c^+\to p\eta$ & 0.96 & $1.11$ & $2.08$ & $-1.93$ & $-1.24$ & $-3.17$ & $1.28\times 10^{-3}$  & $(1.24\pm0.29)10^{-3}$ & $-0.56$ \\
$\Lambda_c^+\to n\pi^+$ & $-1.64$ & 1.15 & $-0.50$ & 3.45 & $-1.57$ & 1.88 & $2.66\times 10^{-4}$ & & $-0.90$ \\
$\Lambda_c^+\to \Lambda K^+$ & $-1.66$ & 0.09 & $-1.57$ & 4.43 & $-0.54$ & 3.70 & $1.06\times 10^{-3}$ & $(6.1\pm1.2)10^{-4}$ & $-0.96$ \\
$\Lambda_c^+\to \Sigma^0 K^+$ & 0 & $-1.48$ & $-1.48$ &0 & 2.30  & 2.30 & $7.18\times 10^{-4}$ & $(5.2\pm0.8)10^{-4}$ & $-0.73$\\
$\Lambda_c^+\to \Sigma^+ K^0$ & 0 & $-2.10$ & $-2.10$ &0 & 3.25  & 3.25 & $1.44\times 10^{-3}$ & $$ & $-0.74$\\
\end{tabular}
\end{ruledtabular}
\end{table}

As for the axial-vector form factors $g^A_{\B'\B}$, they are discussed in Appendix C.

\section{Results and Discussions}

In terms of the decay amplitude  of $\B_i\to\B_f+P$ given in Eq. (\ref{eq:Amp}), its decay rate reads
\be
\Gamma &=& {p_c\over 8\pi}\left\{ {(m_i+m_f)^2-m_P^2\over m_i^2} |A|^2+{(m_i-m_f)^2-m_P^2\over m_i^2} |B|^2\right\}, \non \\
&=& {p_c\over 8\pi}\left\{ {(m_i+m_f)^2-m_P^2\over m_i^2} |A|^2+{4p_c^2\over (m_i+m_f)^2-m_P^2 } |B|^2\right\},
\en
with $p_c$ being the c.m. three-momentum in the rest frame of $\B_i$, and the up-down asymmetry $\alpha$ is given by
\be \label{eq:alpha}
\alpha={2\kappa {\rm Re}(A^*B)\over |A|^2+\kappa^2|B|^2}
\en
with $\kappa=p_c/(E_f+m_f)=\sqrt{(E_f-m_f)/(E_f+m_f)}$. If the parent baryon $\B_i$ is unpolarized,  the produced
baryon $\B_f$ is longitudinally polarized by an amount of $\alpha$.
The predicted $S$- and $P$-wave amplitudes of singly Cabibbo-suppressed decays $\Lambda_c^+\to \Lambda K^+, p\pi^0, p\eta, n\pi^+,\Sigma^0K^+,\Sigma^+ K^0$, their branching fractions and decay asymmetries are shown in Table \ref{tab:S&P}.

\begin{figure}[t]
\begin{center}
\includegraphics[width=0.80\textwidth]{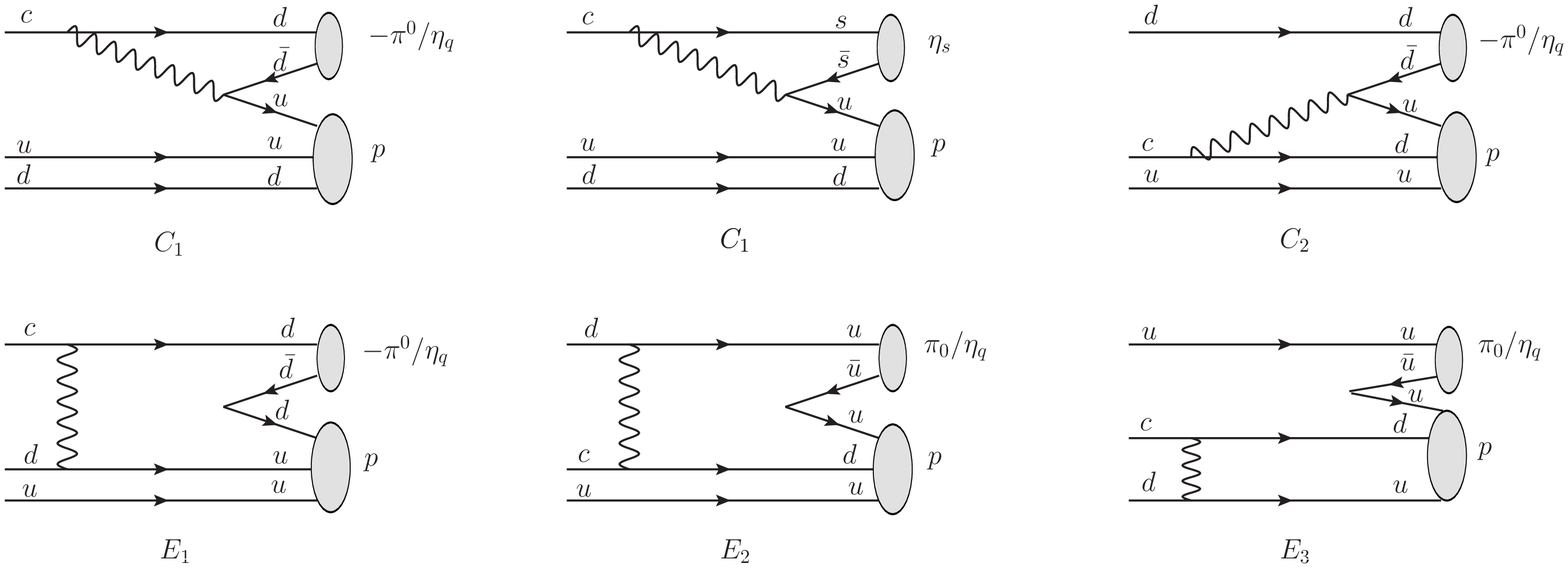}
\vspace{0.5cm}
\caption{Quark diagrams contributing to $\Lambda_c^+\to p\eta$ and $p\pi^0$.} \label{fig:SCS} \end{center}
\end{figure}

We first discuss the two modes $\Lambda_c^+\to p\pi^0$ and $p\eta$.
In the topological quark-diagram approach for charmed baryon decays \cite{Chau}, the relevant quark diagrams for $\Lambda_c^+\to p\eta$, $p\pi^0$ are depicted in Fig. \ref{fig:SCS}. There are two internal $W$-emission diagrams $C_1$ and $C_2$ and three $W$-exchange ones $E_1,E_2$ and $E_3$. Symmetry properties of the baryon wave function are taken into account in the analysis of \cite{Chau}.
Among these diagrams, only $C_1$ is factorizable. Since the CKM matrix elements $V_{cs}V_{us}$ and $V_{cd}V_{ud}$ are similar in magnitude but opposite in sign and since the decay constants $f_\eta^s$ and $f_\eta^q$ also have opposite signs (see Eq. (\ref{eq:decayconst})), it is obvious that factorizable amplitude of $p\eta$ is significantly  larger than $p\pi^0$ in magnitude owing to the constructive interference in the former  (see Table \ref{tab:S&P}). Considering the factorizable contributions alone, we already have $\B(\Lambda_c^+\to p\eta)^{\rm fac}=4.0\times 10^{-4}$, while  $\B(\Lambda_c^+\to p\pi^0)^{\rm fac}=0.93\times 10^{-4}$. We rely on the current-algebra approach to evaluate nonfactorizable $W$-exchange amplitudes,  namely, the commutator terms for the $S$-wave and the current-algebra pole terms for the $P$-wave.

To compute the $\Lambda_c^+\to p\eta$ rate, we have followed \cite{Kroll} to use the decay constant $f_8=1.26 f_\pi$ to get $f_{\eta_8}=f_8\cos(-15.4^\circ)$.
Our prediction  $\B(\Lambda_c^+\to p\eta)=1.28\times 10^{-3}$  is in excellent agreement with the BESIII measurement of $(1.24\pm0.29)\times 10^{-3}$ \cite{BES:peta}.
\footnote{If the hypercharge convention $Y=B+S+C$ is used, we will have $Y(\Lambda_c^+)=2$. In this case, $A^{\rm fac}$ will flip it sign and get a large destructive interference with the $A^{\rm com}$ term. The predicted rate will become very small, $\B(\Lambda_c^+\to p\eta)=1.18\times 10^{-4}$.}
We see from Table \ref{tab:S&P} that the $p\eta$ ($p\pi^0$) mode has a large (small) rate because of a large constructive (destructive) interference between the factorizable and nonfactorizable amplitudes for both $S$- and $P$-waves.

\begin{table}[tbp]
 \caption{Comparison of various theoretical predictions for the branching fractions (in units of $10^{-3}$) of singly Cabibbo-suppressed decays of $\Lambda_c^+$.
 } \label{tab:Comparision}
\centering
\begin{ruledtabular}
\begin{tabular}{l c c c c c c c}
 & Sharma {\it et al.} & Uppal {\it et al.}  & Chen {\it et al.} & Lu {\it et al.} & Geng {\it et al.} & This work & Expt \\
 &  \cite{Sharma} &  \cite{Uppal}  & \cite{Chen:2003} & \cite{Lu} &  \cite{Geng:Lambdac} &  & \cite{BES:peta,PDG} \\
 \hline
$\Lambda_c^+\to p\pi^0$    & 0.2 & 0.1-0.2 & 0.11-0.36 & 0.48 & $0.56\pm0.15$ & 0.08 & $<0.27$ \\
$\Lambda_c^+\to p\eta$ &  0.2\footnotemark[1](1.7)\footnotemark[2] & 0.3 & & & $1.24\pm0.41$  & 1.28 & $1.24\pm0.29$\\
$\Lambda_c^+\to p\eta'$ &  0.4-0.6 & 0.04-0.2 & & & $1.22^{+1.43}_{-0.87}$  &  & $$\\
$\Lambda_c^+\to n\pi^+$    & 0.4 & 0.8-0.9 & 0.10-0.21 & 0.97 &  & 0.27 &  \\
$\Lambda_c^+\to \Lambda K^+$ & 1.4& 1.2 & 0.18-0.39 & & $0.46\pm0.09$ & 1.06 & $0.61\pm0.12$ \\
$\Lambda_c^+\to \Sigma^0 K^+$ & 0.4-0.6 & 0.2-0.8 & & & $0.40\pm0.08$ & 0.72 & $0.52\pm0.08$ \\
$\Lambda_c^+\to \Sigma^+ K^0$ & 0.9-1.2 & 0.4-0.8 & & & $0.80\pm0.16$  & 1.44  &  \\
\end{tabular}
\end{ruledtabular}
\footnotetext[1]{The $P$-wave amplitude of $\Lambda_c^+\to\Xi^0K^+$ is assumed to be positive.}
\footnotetext[2]{The $P$-wave amplitude of $\Lambda_c^+\to\Xi^0K^+$ is assumed to be negative.}
\end{table}

Various other model predictions for the singly Cabibbo-suppressed decays $\Lambda_c^+\to\B+P$ are summarized in Table \ref{tab:Comparision}. Except
for the dynamic calculation in \cite{Uppal} and the consideration of factorizable contributions in \cite{Chen:2003}, all other predictions are based on the SU(3) symmetry argument.   A global fit of the SU(3) amplitudes of $\Lambda_c^+\to\B+P$ to the data of branching fractions of Cabibbo-allowed decays $\Lambda_c^+\to p\overline{K}^0,\Lambda\pi^+,\Sigma^+\pi^0,\Sigma^0\pi^+,\Sigma^+\eta,\Xi^0K^+,$ and singly-Cabibbo-suppressed decays $\Lambda_c^+\to\Lambda K^+,\Sigma^0K^+,p\eta$ in \cite{Geng:Lambdac} yields $\B(\Lambda_c^+\to p\pi^0)=(5.6\pm1.5)\times 10^{-4}$, which is too large compared to the experimental limit of $2.7\times 10^{-4}$ \cite{BES:peta}. Assuming the sextet {\bf 6} dominance over $\overline{\bf 15}$ (i.e. $c_-O_-\gg c_+O_+$), the authors of \cite{Lu} obtained the relation
\footnote{It was also noticed in \cite{Sharma}.}
\be \label{eq:npi}
M(\Lambda_c^+\to n\pi^+)=\sqrt{2}M(\Lambda_c^+\to p\pi^0),
\en
and the sum rule
\be \label{eq:SR}
\B(\Lambda_c^+\to n\pi^+)=\sin^2\theta_C\left[ 3\B(\Lambda_c^+\to\Lambda\pi^+)+
\B(\Lambda_c^+\to\Sigma^0\pi^+)-\B(\Lambda_c^+\to p\overline{K}^0)\right],
\en
derived from the relations \cite{Geng:Lambdac}
\be \label{eq:SU(3)relations}
\sqrt{6}M(\Lambda_c^+\to\Lambda\pi^+)+\sqrt{2}M(\Lambda_c^+\to\Sigma^0\pi^+) &=& 2M(\Lambda_c^+\to p\bar K^0), \non \\
\sqrt{6}M(\Lambda_c^+\to\Lambda\pi^+)-\sqrt{2}M(\Lambda_c^+\to\Sigma^0\pi^+) &=& {2\over \sin\theta_C}M(\Lambda_c^+\to n\pi^+).
\en
The current PDG values for branching fractions \cite{PDG} lead to $\B(\Lambda_c^+\to n\pi^+)\sim 0.97\times 10^{-3}$ and hence $\B(\Lambda_c^+\to p\pi^0)\sim 0.48\times 10^{-3}$. The prediction of the latter is consistent with the SU(3) global fit of \cite{Geng:Lambdac}.
The discrepancy between the SU(3) approach and experiment for $\Lambda_c^+\to p\pi^0$ is ascribed to the SU(3) relations given by Eqs. (\ref{eq:npi}) and (\ref{eq:SR}). First of all, the relation (\ref{eq:npi}) does not hold in the general quark diagram approach owing to the presence of factorizable contributions \cite{Chau}.
Since the factorizable amplitude of $\Lambda_c^+\to n\pi^+$ ($\Lambda_c^+\to p\pi^0$) is governed by the external (internal) $W$-emission, we have (see also Table \ref{tab:S&P})
\be
{M(\Lambda_c^+\to n\pi^+)^{\rm fac}\over M(\Lambda_c^+\to p\pi^0)^{\rm fac}}=-\sqrt{2}\left({a_1\over a_2}\right)\approx 2.8\sqrt{2}.
\en
Hence, the factorizable amplitudes alone strongly violate the SU(3) relation (\ref{eq:npi}). If we just consider the operator $c_-O_-^d$ alone, it is easily seen that naive factorization leads to $a_1={2\over 3}c_-$ and $a_2=-{2\over 3}c_-$, and hence $M(\Lambda_c^+\to n\pi^+)^{\rm fac}=\sqrt{2}M(\Lambda_c^+\to p\pi^0)^{\rm fac}$. However, in reality $a_1\sim 1.26 \gg |a_2|\sim 0.45$.
Since the matrix element $a_{\B'\B}$ is governed by the operator $O_-$, it is clear that the relation (\ref{eq:npi}) should be  respected by $A^{\rm com}$ and $B^{\rm ca}$ terms, but not by $A^{\rm fac,tot}$ and $B^{\rm fac,tot}$ (see Table \ref{tab:S&P}).
By the same token, the first line of Eq. (\ref{eq:SU(3)relations}) does not hold as the factorizable amplitudes of $\Lambda_c^+\to\Lambda\pi^+$ and $\Lambda_c^+\to p \bar K^0$ are of different types, governed by $a_1$ and $a_2$, respectively.  Hence, we conclude that the rates of $n\pi^+$ and $p\pi^0$ cannot be extracted from experiment through the invalid SU(3) relations (\ref{eq:SR}) and (\ref{eq:npi}). In our work, both $n\pi^+$ and $p\pi^0$ are suppressed owing to the destructive interference between factorizable and nonfactorizable terms.
Experimentally, the Cabibbo-allowed decay $\Lambda_c^+\to nK_S \pi^+$ involving a neutron was observed by BESIII recently \cite{BES:nKpi}. It is conceivable that the Cabibbo-suppressed mode $\Lambda_c^+\to n\pi^+$ can be reached in the near future.

Only factorizable contributions to $\Lambda_c^+\to n\pi^+$ and $p\pi^0$ were considered in \cite{Chen:2003}. In the naive factorization with $N_c^{\rm eff}=3$, the branching ratio of $\Lambda_c^+\to p\pi^0$ of order $10^{-6}$ is smaller than that of $\Lambda_c^+\to n\pi^+$ by a factor of order 50. It was argued in \cite{Chen:2003} that final-state rescattering effects through $\Lambda_c^+\to \{n\pi^+,n\rho^+,\Lambda K^+,\Lambda K^{+*}\}\to p\pi^0$  will enhance the former so that
$\B(\Lambda_c^+\to p\pi^0)\gsim \B(\Lambda_c^+\to n\pi^+)$
(see Tables 2 and 3 of \cite{Chen:2003}).  We would like to make two remarks: (i) In order to enhance the rate of $p\pi^0$ to the order of $10^{-4}$, a common wisdom is that the branching fraction of the intermediate states, e.g. $\Lambda_c^+\to n\rho^+,\Lambda K^+$, should be at least two orders of magnitude larger than  $10^{-4}$ \cite{Cheng:FSI}. (ii) We find that even in the absence of final-state rescattering, the nonfactorizable contributions denoted by $A^{\rm com}$ and $B^{\rm ca}$ in Table \ref{tab:S&P}, which were neglected in \cite{Chen:2003}, will yield $\B(\Lambda_c^+\to p\pi^0)^{\rm nf}=3.3\times 10^{-4}$. Therefore, it is mandatory to take into account the nonfactorizable contributions from internal $W$-emission (denoted by $C_2$ in Fig. \ref{fig:SCS}) and $W$-exchange ($E_1,E_2,E_3$) in the study.

As for $\Lambda_c^+\to\Sigma K$ decays, we see from Eqs. (\ref{eq:a}) and (\ref{eq:m.e.}) that $a_{\Xi_c^+\Sigma^+}\cong -\sqrt{2}a_{\Xi_c^0\Sigma^0}$ due to the smallness of the bag integrals $X_1^{d,s}$ compared to $X_2^s$ (see Eq. (\ref{eq:Xnum})). It follows from Eq. (\ref{eq:Swave}) that $A(\Lambda_c^+\to \Sigma^+ K^0)\cong \sqrt{2}A(\Lambda_c^+\to\Sigma^0 K^+)$.
Moreover, the relations $g^A_{\Sigma^+ p}=-\sqrt{2}g^A_{\Sigma^0 p}$ and  $g_{\Xi_c^{'0}\Lambda _c}^A=-g_{\Xi_c^{'+}\Lambda _c}^A$ (c.f. Appendix C) and the identity  $a_{\Sigma^0\Xi_c^{'0}}=a_{\Sigma^+\Xi_c^{'+}}$ for matrix elements also lead to $B(\Lambda_c^+\to \Sigma^+ K^0)= \sqrt{2}B(\Lambda_c^+\to\Sigma^0 K^+)$, see Eq. (\ref{eq:Pwave}). It is thus expected that
\be
\Gamma(\Lambda_c^+\to \Sigma^+ K^0)\cong 2\Gamma(\Lambda_c^+\to\Sigma^0 K^+),
\en
and identical decay asymmetries in both channels.

We did not consider the decay mode $\Lambda_c^+\to p\eta'$ as the evaluation of its nonfactorizable amplitude is beyond the current-algebra framework. Nevertheless, We find from Eq. (\ref{eq:facamp}) that $\B(\Lambda_c^+\to p\eta')^{\rm fac}=0.9\times 10^{-4}$ due to the factorizable effect alone. Notice that it has been claimed in \cite{Geng:Lambdac} that its branching fraction is as large as $\Lambda_c^+\to p\eta$, namely,  $\B(\Lambda_c^+\to p\eta')=(1.22^{+1.43}_{-0.87})\times 10^{-3}$.

\begin{table}[t]
\caption{Same as Table \ref{tab:Comparision} except for decay asymmetries. The $P$-wave amplitude of $\Lambda_c^+\to\Xi^0K^+$ is assumed to be positive (negative) in  case $a$ ($b$) \cite{Sharma}, while $|\psi(0)|^2$ scale violation is (not) taken into account  in case $c$ ($d$) \cite{Uppal}.
}
\label{tab:alpha}
\begin{center}
\begin{tabular}{| l c c c |} \hline
 & Sharma {\it et al.} \cite{Sharma}~~ & Uppal {\it et al.} \cite{Uppal} &  ~~This work~~ \\
 \hline
~$\Lambda_c^+\to p\pi^0$    & 0.05\footnotemark[1](0.05)\footnotemark[2] & 0.82\footnotemark[3](0.85)\footnotemark[4] & $-0.95$ \\
~$\Lambda_c^+\to p\eta$ &  $-0.74$($-0.69$) & ~$-1.00(-0.79)$ & $-0.56$ \\
~$\Lambda_c^+\to p\eta'$ & $-0.97(-0.99)$ & 0.87(0.87) & \\
~$\Lambda_c^+\to n\pi^+$    & 0.05(0.05) & $-0.13$(0.67) & $-0.90$   \\
~$\Lambda_c^+\to \Lambda K^+$ & $-0.54(0.97)$ & $-0.99(-0.99)$  & $-0.96$ \\
~$\Lambda_c^+\to \Sigma^0 K^+$ & 0.68($-0.98)$ & $-0.80(-0.80)$ & $-0.73$ \\
~$\Lambda_c^+\to \Sigma^+ K^0$ & 0.68($-0.98)$ & $-0.80(-0.80)$ & $-0.74$ \\
\hline
\end{tabular}
\end{center}
\end{table}

For the decay $\Lambda_c^+\to\Lambda K^+$, if we follow \cite{Gutsche} to use the form factors
$f_1^{\Lambda_c\Lambda}(0)=0.511$ and $g_1^{\Lambda_c\Lambda}(0)=0.466$, we will obtain $\B(\Lambda_c^+\to\Lambda K^+)\sim 1.9\times 10^{-3}$, which is too large by a factor of three compared to experiment. The same is also true for the Cabibbo-allowed decay $\Lambda_c^+\to \Lambda \pi^+$. Using the same set of $\Lambda_c-\Lambda$ transition form factors, we find $\B(\Lambda_c^+\to \Lambda \pi^+)=2.4\%$, while it is $(1.30\pm0.07)\%$ experimentally \cite{PDG}. Nevertheless, the predicted ratio $R\equiv \Gamma(\Lambda K^+)/\Gamma(\Lambda \pi^+)=0.078$, which is close to $(\sin^2\theta_C f_K)^2$, is smaller than the BaBar measurement of $R=0.044\pm0.005$ \cite{BaBar:LambdaK}, but consistent with the Belle's value of $0.074\pm0.016$ \cite{Belle:LambdaK}. An average of $R=0.047\pm0.009$ is quoted by PDG \cite{PDG}. In Table \ref{tab:S&P} we use the form factors $f_1^{\Lambda_c\Lambda}(0)=0.406$ and $g_1^{\Lambda_c\Lambda}(0)=0.370$ fitted to $\Lambda_c^+\to\Lambda \pi^+$ to predict $\Lambda_c^+\to\Lambda K^+$.

Finally, other model calculations for the up-down decay asymmetry are collected in Table \ref{tab:alpha} for comparison.
The predicted decay asymmetries under current algebra for the singly Cabibbo-suppressed modes are always negative and range from $-0.56$ to $-0.96$.
The SU(3) approach usually cannot give definite predictions without further assumptions.

\section{Conclusions}
We have studied singly Cabibbo-suppressed two-body hadronic decays of the charmed baryon $\Lambda_c^+$.    We use the measured rate of  $\Lambda_c^+\to p\phi$ to fix the effective Wilson coefficient $a_2$ for naive color-suppressed modes and the effective number of color $N_c^{\rm eff}$. We rely on the current-algebra method to evaluate $W$-exchange and nonfactorizable internal $W$-emission amplitudes,  that is,  the commutator terms for the $S$-wave and the  pole terms for the $P$-wave.
Our prediction for $\Lambda_c^+\to p\eta$  is in excellent agreement with the BESIII measurement. The $p\eta$ ($p\pi^0$) mode has a large (small) rate because of a large constructive (destructive) interference between the factorizable and nonfactorizable amplitudes for both $S$- and $P$-waves.
Some of the SU(3) relations such as  $M(\Lambda_c^+\to n\pi^+)=\sqrt{2}M(\Lambda_c^+\to p\pi^0)$  and Eq. (\ref{eq:SR}) derived under the assumption of sextet dominance are not valid for decays with factorizable contributions. Sextet dominance is justified for nonfactorizable terms as the baryon matrix elements $a_{\B'\B}$ are governed by the four quark operator $O_-$, but not for factorizable amplitudes as both $O_-$ and $O_+$ operators contribute.
Our calculation indicates that  the branching fraction of $\Lambda_c^+\to n\pi^+$  is about 3.5 times larger than that of $\Lambda_c^+\to p\pi^0$.  Decay asymmetries are found to be negative  for all singly Cabibbo-suppressed modes and range from $-0.56$ to $-0.96$.

\vskip 2.0cm \acknowledgments
This research was supported in part by the Ministry of Science and Technology of R.O.C. under Grant No. 106-2112-M-001-015. F. Xu is supported by NSFC under Grant No. 11605076 as well as the Fundamental Research
Funds for the Central Universities in China under the Grant No. 21616309.

\appendix

\section{The decay $\Lambda_c^+\to p\,\phi$}

The decay $\Lambda_c^+\to p\phi$ proceeds only through the internal $W$-emission  governed by
\be
M(\Lambda_c^+\to p\phi)= {G_F\over \sqrt{2}}V_{cs}V_{us}a_2\la \phi|(\bar ss)|0\ra\la p|(\bar u c)|\Lambda_c^+\ra.
\en
Given the general amplitude
\be
  \bar{u}_f(p_f)\varepsilon^{*\mu}[A_1
\gamma_\mu\gamma_5+A_2p_{f\mu}\gamma_5+B_1\gamma_\mu+B_2 p_{f\mu} ] u_i(p_i),
\en
for the decay $\B_i(1/2^+)\to \B_f(1/2^+)+V$,  we find
\be
A_1 &=& -a_2\, hf_\phi m_\phi\left[g_1^{\Lambda_c p}(m_\phi^2)-g_2^{\Lambda_c p}(m^2_\phi)(m_{\Lambda_c}-m_p)/m_{\Lambda_c} \right],   \non \\
A_2 &=& 2a_2\,hf_\phi m_\phi g_2^{\Lambda_c p}(m^2_\phi)/m_{\Lambda_c},   \non \\
B_1 &=& a_2\,hf_\phi m_\phi \left[f_1^{\Lambda_c p}(m_\phi^2)+f_2^{\Lambda_c p}(m^2_\phi)(m_{\Lambda_c}+m_p)/m_{\Lambda_c} \right],   \\
B_2 &=& 2a_2\,hf_\phi m_\phi f_2^{\Lambda_c p}(m^2_\phi)/m_{\Lambda_c},  \non
\en
with $h=G_F V_{cs}V_{us}/\sqrt{2}$, where use of
the decay constant $f_\phi$ defined by $\la \phi|(\bar ss)|0\ra=f_\phi m_\phi\varepsilon_\mu^*$ and form factors defined in analog to Eq. (\ref{eq:FF}) has been made.  The partial decay rate and decay asymmetry then read
\footnote{The formula for $\Gamma(1/2^+\to 1/2^++V)$ given in Eq. (\ref{eq:Vrate}) in terms of partial wave amplitudes was originally derived in \cite{Pak} and has been widely used in the literature. However, the original expression for $\Gamma$ is too small by a factor of 2. We would like to thank Hong-Wei Ke for pointing this out to us. Later, we learned that the correct expression of $\Gamma$ was also obtained by Wang, Yu and Zhao in the spring of 2017 (see Eq. (57) of \cite{Yu}).
It should be stressed that both
$\Gamma$ and $\alpha$ can be expressed in terms of the helicity amplitudes defined by  $h_{\lambda_f,\lambda_V;\lambda_i}=\,\la \B_f(\lambda_f)V(\lambda_V)|H_W|\B_i(\lambda_i)\ra$ with $\lambda_i=\lambda_f-\lambda_V$ \cite{Korner} which yield the same results as Eq. (\ref{eq:Vrate}). Hence, the partial-wave method and the helicity-amplitude method are equivalent.}
\be \label{eq:Vrate}
\Gamma(\Lambda_c^+\to p\phi) &=& {p_c\over 4\pi}\,{E_p+m_p\over m_{\Lambda_c}}\left[
2(|S|^2+|P_2|^2)+{E^2_\phi\over m_\phi^2}(|S+D|^2+|P_1|^2)\right], \non \\
\alpha(\Lambda_c^+\to p\phi) &=& {4m^2_\phi{\rm Re}(S^*P_2)+2E^2_\phi{\rm Re}(S+D)^*
P_1\over 2m_\phi^2(|S|^2+|P_2|^2)+E^2_\phi(|S+D|^2+|P_1|^2)},
\en
with the $S$-, $P$- and $D$-waves given by \cite{Pak,Cheng:1997}
\be
S &=& -A_1,   \non \\
P_1 &=& -{p_c\over E_\phi}\left( {m_{\Lambda_c}+m_p\over E_p+m_p}B_1+m_{\Lambda_c}B_2\right), \non \\
P_2 &=& {p_c\over E_p+m_p}B_1,   \\
D &=& -{p_c^2\over E_\phi(E_p+m_p)}\,(A_1-m_{\Lambda_c}A_2). \non
\en
Using the data $\B(\Lambda_c^+\to p\phi)=(1.04\pm0.21)\times 10^{-3}$ \cite{BES:peta}, we obtain $|a_2|=0.45\pm0.03$ and hence $N_c^{\rm eff}\approx 7$ for $c_1 = 1.346$ and $c_2 = -0.636$ and $f_\phi=215$ MeV. This leads to $a_1=1.26\pm 0.01$\,.

\section{Baryon matrix elements in the bag model}
For the evaluation of the baryon  matrix element of $O$ in the MIT bag model \cite{MIT}, see \cite{Cheng:1992}. Consider the four-quark operator $O=(\bar q_1q_3)(\bar q_2q_4)$. This operator can be written as
$O=6(\bar q_1q_3)_1(\bar q_2q_4)_2$, where the superscript $i$ indicates that the quark operator acts only on the $i$th quark in the baryon wave function.
In the bag model the parity-conserving matrix elements have the expression \cite{Cheng:1992}
\be
\int r^2dr\la q_1q_2|(\bar q_1q_3)_1(\bar q_2q_4)_2|q_3q_4\ra &=& (-X_1+X_2)-{1\over 3}(X_1+3X_2){\boldsymbol\sigma}_1\cdot {\boldsymbol\sigma}_2, \non \\
\int r^2dr\la q_1q_2|(\bar q_1q_4)_1(\bar q_2q_3)_2|q_3q_4\ra &=& (X_1+X_2)-{1\over 3}(-X_1+3X_2){\boldsymbol\sigma}_1\cdot{\boldsymbol\sigma}_2,
\en
with
\be
X_1 &=& \int_0^R r^2dr(u_1v_2-v_1u_2)(u_3v_4-v_3u_4), \non \\
X_2 &=& \int_0^R r^2dr(u_1u_2+v_1v_2)(u_3u_4+v_3v_4),
\en
where $R$ is the radius of the bag and $u(r), v(r)$ are the large and small components of the quark wave function, respectively, defined by
\be \label{eq:uv}
\psi=
\left(
\begin{array}{c}
iu(r)\chi \\
v(r){\boldsymbol\sigma}\cdot{\bf \hat{r}}\chi \\
\end{array}
\right).
\en

As an example for illustration, we consider the matrix element $a_{p\Lambda_c}$
given by
\be
a_{p\Lambda_c} \equiv  \la p|H_{\rm eff}^{\rm PC}|\Lambda_c^+\ra =
{G_F\over2\sqrt{2}}V_{cd}V_{ud}c_-\la p|O_-^d|\Lambda_c^+\ra.
\en
Applying the relation
\be
{\boldsymbol\sigma}_1\cdot{\boldsymbol\sigma}_2={1\over 2}(\sigma_{1+}\sigma_{2-}+\sigma_{1-}\sigma_{2+})+\sigma_{1z}\sigma_{2z},
\en
and the wave functions (see Appendix D)
\be
\Lambda_c^+ &=& -{1\over \sqrt{12}}\left[u^\uparrow d^\downarrow c^\uparrow-u^\dw d^\up c^\up-d^\up u^\dw c^\up+d^\dw u^\up c^\up+(13)+(23)\right], \non \\
 p&=& {1\over \sqrt{18}}\left[2u^\uparrow u^\up d^\dw -u^\up u^\dw d^\up- u^\dw u^\up d^\up+(13)+(23)\right],
\en
with obvious notation for permutation of quarks,
it is straightforward to show that
\be \label{eq:Y2Y1}
\la p|(\bar dc)(\bar ud)|\Lambda_c^+\ra
&=& 6(4\pi)\la p|b^\dagger_{1d}b_{1c}b^\dagger_{2u}b_{2d}\left[-X_1^d+X_2^d-{1\over 3}(X_1^d+3X_2^d){\boldsymbol\sigma}_1\cdot {\boldsymbol\sigma}_2\right]|\Lambda_c^+\ra \non \\
&=& \sqrt{2\over 3}(X_1^d+3X_2^d)(4\pi), \non \\
\la p|(\bar dd)(\bar uc)|\Lambda_c^+\ra
&=& 6(4\pi)\la p|b^\dagger_{1d}b_{1d}b^\dagger_{2u}b_{2c}\left[X_1^d+X_2^d-{1\over 3}(-X_1^d+3X_2^d){\boldsymbol\sigma}_1\cdot {\boldsymbol\sigma}_2 \right]|\Lambda_c^+\ra \non \\
&=& -\sqrt{2\over 3}(X_1^d+3X_2^d)(4\pi),
\en
with
\be \label{eq:baginteg}
X_1^q &=& \int_0^R r^2 dr(u_qv_u-v_qu_u)(u_qv_c-v_qu_c), \non \\
X_2^q &=& \int_0^R r^2 dr(u_qu_u+v_qv_u)(u_qu_c+v_qv_c)
\en
for $q=d,s$. Hence,
\be
\la p|O_-^d|\Lambda_c^+\ra ={4\over \sqrt{6}}(X_1^d+3X_2^d)(4\pi).
\en
Numerically, we obtain
\be \label{eq:Xnum}
X_1^d=0, \quad X_2^d=1.60\times 10^{-4}, \quad X_1^s=2.60\times 10^{-6}, \quad X_2^s=1.96\times 10^{-4},
\en
where we have employed the following bag parameters
\be
m_u=m_d=0, \quad m_s=0.279~{\rm GeV}, \quad m_c=1.551~{\rm GeV}, \quad R=5~{\rm GeV}^{-1}.
\en

\section{Axial-vector form factor $g_{\B'\B}^A$}
To evaluate the $S$- and $P$-wave amplitudes in the pole model, one needs to know the strong couplings $g_{\B^*\B P}$ and $g_{\B'\B P}$ in Eq. (\ref{eq:amppole}).  In the approach of current algebra, the $\B'\B P$ coupling is related to $g_{\B'\B}^A$, the axial-vector form factor at $q^2=0$ through the Goldberger-Treiman relation given in Eq. (\ref{eq:GT}). In the bag model the axial form factor in the static limit has the expression \cite{Cheng:1993}
\be
g_{\B'\B}^A=\la \B' \up|b^\dagger_{q1}b_{q2}\sigma_z|\B \up\ra\int d^3r(u_{q1}u_{q2}-{1\over 3}v_{q1}v_{q2}).
\en
Here we show those results relevant to the present work:
\be \label{eq:gAoctet}
g_{np}^A=\sqrt{2}g_{pp}^{A(\pi^0)}=5\sqrt{2}g_{pp}^{A(\eta_8)}={5\over 3}(4\pi Z_1),\quad g_{\Lambda p}^A=-{\sqrt{6}\over 2}(4\pi Z_2),\quad g_{\Sigma^+ p}^A=\sqrt{2}g_{\Sigma^0 p}^A={1\over 3}(4\pi Z_2),\quad
\en
for octet baryons,
\be
g_{\Sigma_c^+\Lambda_c}^{A(\pi^0)}={1\over\sqrt{2}}g_{\Sigma_c^0\Lambda_c}^A={1\over \sqrt{3}}(4\pi Z_1), \quad g_{\Sigma_c^+\Lambda_c}^{A(\eta_8)}=0, \quad g_{\Xi_c^{'0}\Lambda _c}^A=-g_{\Xi_c^{'+}\Lambda _c}^A={1\over \sqrt{3}}(4\pi Z_2),
\en
and
\be
g_{\B_{\bar 3}\B_{\bar 3}}^A=0
\en
for charmed baryons, where
\be
Z_1=\int r^2dr (u_u^2-{1\over 3}v_u^2),  \qquad
Z_2=\int r^2dr (u_u u_s-{1\over 3}v_u v_s),
\en
and $\B_{\bar 3}$ is an antitriplet heavy baryon, $\Lambda_c^+,\Xi_c^0$ and $\Xi_c^+$. To compute the form factors $g_{pp}^{A(\pi^0)}$ and $g_{\Sigma_c^+\Lambda_c}^{A(\pi^0)}$, we notice the axial-vector current for $P^3=\pi^0$ is ${1\over 2}(\bar u\gamma_\mu\gamma_5 u-\bar d\gamma_\mu\gamma_5 d)$, and likewise for the form factors $g_{pp}^{A(\eta_8)}$ and $g_{\Sigma_c^+\Lambda_c}^{A(\eta_8)}$,
Although the quark model leads to $g_{\B_{\bar 3}\B_{\bar 3}}^A=0$, it is indeed a model-independent result in the heavy quark limit. In the limit of $m_Q\to\infty$,  the diquark of the antitriplet baryon $\B_{\bar 3}$ is a scalar diquark with $J^P=0^+$. Therefore, the diquark transition is $0^+\to 0^+ + 0^-$ for  $\B_{\bar 3}\to \B_{\bar 3}+P$ and it does not conserve parity.

Numerically, we obtain $4\pi Z_1=0.65$ and $4\pi Z_2=0.71$. Using the      Goldberger-Treiman relation and the results of (\ref{eq:gAoctet}), we find that the strong couplings for octet baryons are consistent with those in \cite{Khanna}.

\section{Baryon wave functions}
In the present work, we use the following wave functions for octet and charmed baryons with $S_z=1/2$:
\be
p &=& {1\over\sqrt{3}}\left[ uud\chi_{_S}+(13)+(23)\right], \qquad \qquad \quad ~~n = -{1\over\sqrt{3}}\left[ ddu\chi_{_S}+(13)+(23)\right], \non \\
\Sigma^+ &=& -{1\over\sqrt{3}}\left[ uus\chi_{_S}+(13)+(23)\right], \qquad\qquad \quad
\Sigma^0 = {1\over\sqrt{6}}\left[ (uds+dus)\chi_{_S}+(13)+(23)\right], \non \\
 \Xi^0&=& {1\over\sqrt{3}}\left[ ssu\chi_{_S}+(13)+(23)\right], \qquad \qquad\quad ~~ \Xi^- = {1\over\sqrt{3}}\left[ ssd\chi_{_S}+(13)+(23)\right],  \non \\
\Lambda &=& -{1\over\sqrt{6}}\left[ (uds-dus)\chi_{_A}+(13)+(23)\right], \quad
\Lambda_c^+ = -{1\over\sqrt{6}}\left[ (udc-duc)\chi_{_A}+(13)+(23)\right],  \non \\
\Sigma_c^+ &=& {1\over\sqrt{6}}\left[ (udc+duc)\chi_{_S}+(13)+(23)\right], \qquad
\Sigma_c^0 = {1\over\sqrt{3}}\left[ ddc\chi_{_S}+(13)+(23)\right],  \\
\Xi_c^+ &=& {1\over\sqrt{6}}\left[ (usc-suc)\chi_{_A}+(13)+(23)\right], \qquad
\Xi_c^0 = {1\over\sqrt{6}}\left[ (dsc-sdc)\chi_{_A}+(13)+(23)\right], \non \\
\Xi_c^{'+} &=& {1\over\sqrt{6}}\left[ (usc+suc)\chi_{_S}+(13)+(23)\right], \qquad
\Xi_c^{'0} = {1\over\sqrt{6}}\left[ (dsc+sdc)\chi_{_S}+(13)+(23)\right], \non
\en
where $abc\chi_{_S}=(2a^\up b^\up c^\dw-a^\up b^\dw c^\up-a^\dw b^\up c^\up)/\sqrt{6}$ and $abc\chi_{_A}=(a^\up b^\dw c^\up-a^\dw b^\up c^\up)/\sqrt{2}$.

\newcommand{\bi}{\bibitem}

\end{document}